\documentclass[fleqn,twoside]{article}
\usepackage{espcrc2}

\usepackage{graphicx}

\newcommand{\be}{\begin{equation}}
\newcommand{\ee}{\end{equation}}

\newcommand{\chiPT}{$\chi$PT }
\newcommand{\FRRchiPT}{FRR$\chi$PT }

\title{
\vspace{-2.6cm}
\hfill \rm \null \hfill
\hbox{\normalsize ADP-03-148/T589} \\
\vspace{-2mm}
\hfill \hbox{\normalsize DESY 03-211} \\
\vspace{1.63cm}
The strangeness magnetic moment of the nucleon from FLIC fermions}

\author{D.~B.~Leinweber\address[CSSM]{Special Research Center for the
                       Subatomic Structure of Matter, and		\\
                       Department of Physics, University of Adelaide
                       Adelaide SA 5005  Australia},
        S.~Boinepalli\addressmark[CSSM], 
        A.~W.~Thomas\addressmark[CSSM], 
        A.~G.~Williams\addressmark[CSSM],
        R.~D.~Young\addressmark[CSSM],
        J.~M.~Zanotti\addressmark[CSSM]\address[DESY]{John von
          Neumann-Institut f\"ur Computing  NIC, \\
          Deutsches Elektronen-Synchrotron DESY, D-15738 Zeuthen,
                       Germany} and
        J.~B.~Zhang\addressmark[CSSM]
}

\begin{document}

\begin{abstract}
By imposing the constraints of charge symmetry we show that the
strangeness magnetic moment of the nucleon can be expressed in terms
of empirical magnetic moments and ratios of valence quark magnetic
moments.  The latter are determined using modern chiral extrapolation
techniques and recent low mass lattice QCD simulations of the
individual quark contributions to the magnetic moments of the nucleon
octet.  The result is a precise determination of $G_M^s$, namely
$-0.043 \pm 0.026\ \mu_N$, which is consistent with the latest
experimental measurements.
\end{abstract}


\maketitle

\section{INTRODUCTION}

There is currently enormous interest in the determination of the
strangeness content of the nucleon. It is clearly crucial to our
understanding of QCD to determine precisely the role played by
heavier, non-valence flavors. On the experimental side there have been
tremendous advances in the ability to measure parity violation in
electron scattering at the level of $10^{-7}$ and new results on
strangeness in the nucleon have been reported recently from JLab
(HAPPEX) \cite{Aniol:2000at} and MIT-Bates (SAMPLE)
\cite{Hasty:2001ep}. In the near future we can expect even more
precise results from the A4 experiment at Mainz as well as G0 and
HAPPEX2 at JLab.

In contrast, the theoretical situation is somewhat confused with the
predictions of various quark models covering an enormous range. Direct
calculations within lattice QCD have not yet clarified the situation,
with values for $G_M^s$ ranging from $-0.28 \pm 0.10$
\cite{Mathur:2000cf} to $+0.05 \pm 0.06$ \cite{Lewis:2002ix}.  We take
a different approach, building on the improvements in both lattice
actions and computer speed which have enabled quenched QCD (QQCD)
simulations of magnetic moments at pion masses as low as 0.3--0.4 GeV
and on the developments of modern chiral extrapolation techniques
which allow one to rigorously ensure the model independent constraints
of chiral symmetry.  Using these techniques we determine the ratios of
the $u$-quark contribution to the magnetic moment of the physical
proton to that in the $\Sigma^+$ and of the $u$ quark in the physical
neutron to that in the $\Xi^0$.  From these ratios, experimental data
on the octet moments and charge symmetry, which is typically satisfied
at the level of 1\% or better \cite{ChargeSymm}, we deduce a new
theoretical value for $G_M^s$ which is extremely precise -- setting a
tremendous challenge for the next generation of parity violation
experiments.
 
\section{CHARGE SYMMETRY}

\begin{figure}[tbp]
\vspace*{0.2cm}
{\includegraphics[height=3.3cm,angle=90]{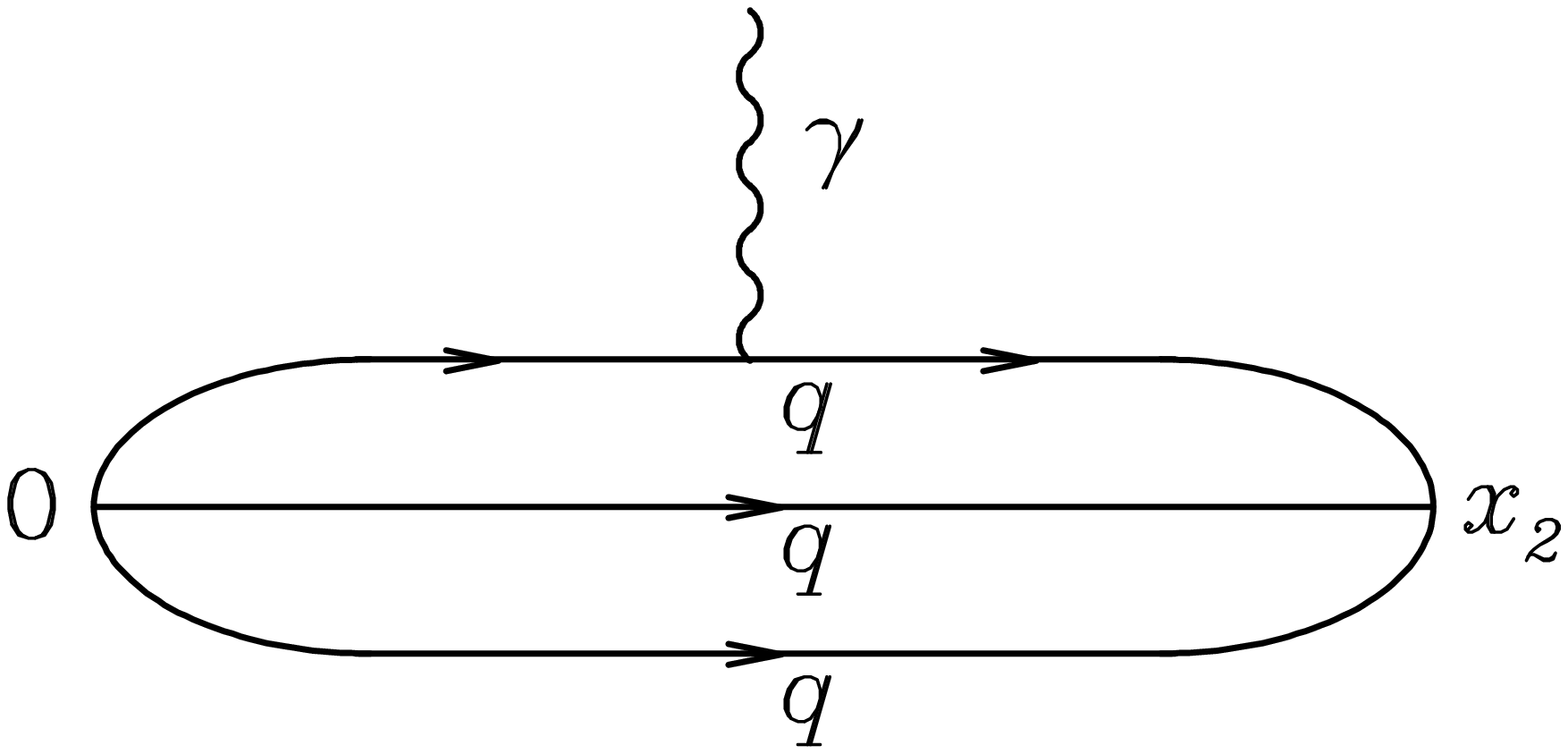} \hfill
 \includegraphics[height=3.3cm,angle=90]{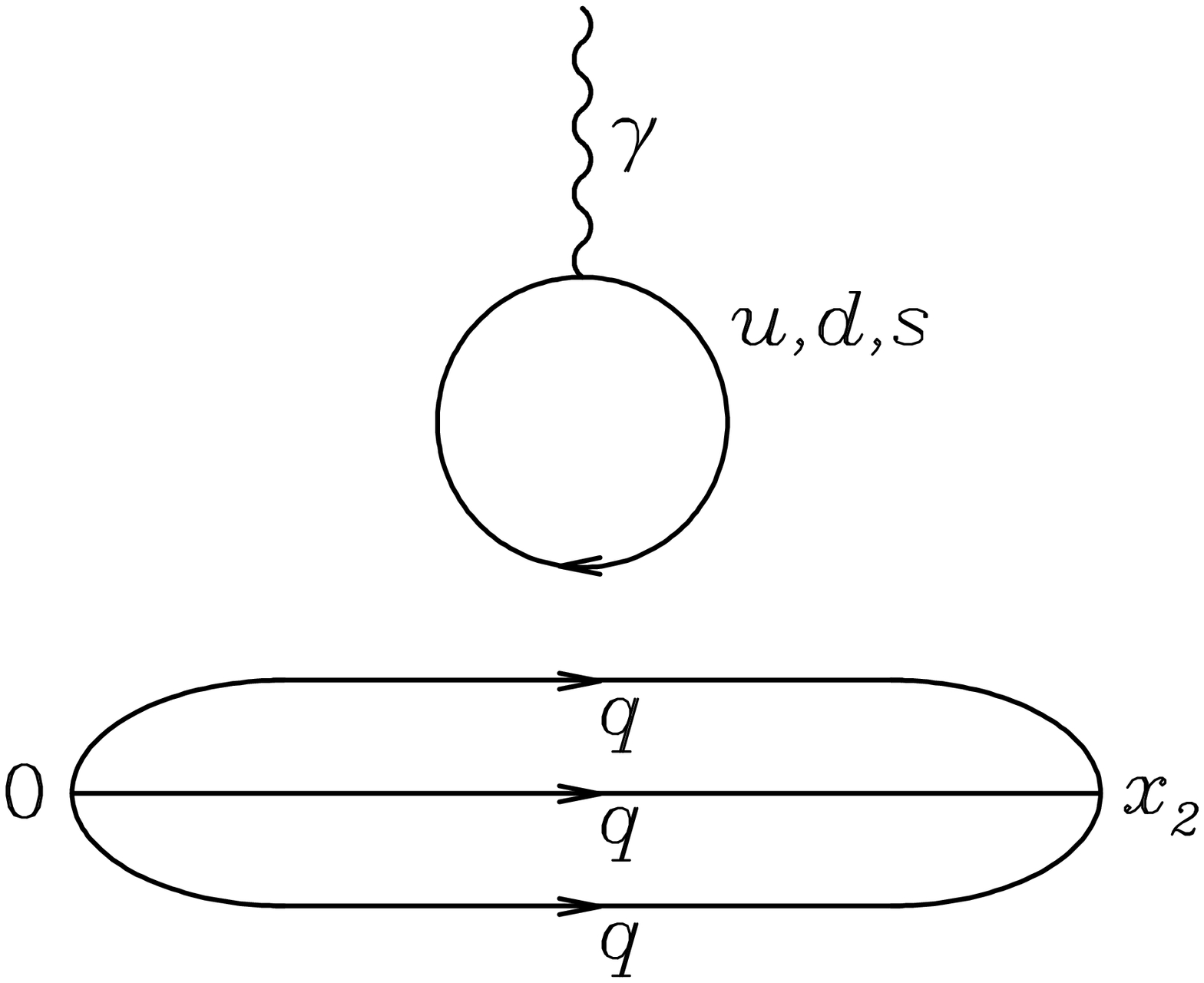}}
\vspace{-0.7cm}
\caption{Diagrams illustrating the two topologically different
insertions of the current within the framework of lattice QCD.  
These skeleton diagrams for the connected
(left) and disconnected (right) current insertions may be dressed by
an arbitrary number of gluons and quark loops.}
\label{topology}
\end{figure}

An examination of the symmetries manifest in the QCD path integral for
current matrix elements reveals various relationships among the quark
contributions \cite{Leinweber:1996ie}. The magnetic moment of the
proton, is extracted from the three-point function, where an operator
exciting the proton from the QCD vacuum is followed by the
electromagnetic current, which in turn is followed by an operator
annihilating the proton back to the QCD vacuum.  In calculating this
three point function, one encounters two topologically distinct ways
of performing the electromagnetic current insertion.
Figure~\ref{topology} displays skeleton diagrams for these two
possible insertions (with Euclidean time increasing to the right).  In
full QCD these diagrams incorporate an arbitrary number of gluons and
quark loops.  The left-hand diagram illustrates the connected
insertion of the current to one of the ``valence'' quarks of the
baryon.  In the right-hand diagram the external field produces a $q \,
\overline q$ pair which in turn interacts with the valence quarks of
the baryon via gluons. It is important to realize that within the
lattice QCD calculation of the loop diagram on the right side of
Fig.~\ref{topology} there is no anti-symmetrization (Pauli blocking)
of the quark in the loop with the valence quarks. For this reason, in
general only the sum of the two processes in Fig.~\ref{topology} is
physical.
%

Under the assumption of charge symmetry, the three-point correlation
functions for octet baryons leads to the following equalities for
electromagnetic current matrix elements \cite{Leinweber:1996ie}:
\begin{eqnarray}
p &=& e_u\, u^p + e_d\, d^p + O_N  \, , \nonumber \\
n &=& e_d\, u^p + e_u\, d^p + O_N  \, , \nonumber \\
\Sigma^+ &=& e_u\, u^{\Sigma} + e_s\, s^\Sigma + O_\Sigma  \, , \nonumber \\
\Sigma^- &=& e_d\, u^{\Sigma} + e_s\, s^\Sigma + O_\Sigma  \, , \nonumber  \\
\Xi^0 &=& e_s\, s^\Xi + e_u\, u^{\Xi} + O_\Xi  \, ,  \nonumber \\
\Xi^- &=& e_s\, s^\Xi + e_d\, u^{\Xi} + O_\Xi  \, .
\label{equalities}
\end{eqnarray}
Here, all quantities refer to magnetic moments, so for example $p$ and
$\Xi^-$ are the physical magnetic moments of the proton and $\Xi^-$,
respectively. The valence $u$-quark magnetic moment in the proton,
corresponding to the left-hand side of Fig.~\ref{topology}, is denoted
$u^p$. We emphasize that charge symmetry has been used to replace the
$d$-quark contribution in the neutron by $u^p$, $d$ in the $\Sigma^-$
by $u$ in the $\Sigma^+$ ( $u^\Sigma$), and so on.  Finally, $O$
denotes the total contribution from quark-loops -- i.e., the
contribution shown on the right-hand side of Fig.~\ref{topology}. The
labels on quark magnetic moments allow for the environment sensitivity
implicit in the three-point function \cite{Leinweber:1996ie}.  For
example, the three-point function for the $\Sigma^+$ is the same as
that for the proton, except that the $d$ is replaced by the somewhat
heavier $s$-quark.  Hence, the $u$-quark propagators in the $\Sigma^+$
are multiplied by an $s$-quark propagator, whereas in the proton they
are multiplied by a $d$-quark propagator.  The different mass of the
neighboring quark gives rise to an environment sensitivity in the
$u$-quark contributions to observables, which means that the naive
expectations of the constituent quark model $u^p/u^{\Sigma} =
u^n/u^{\Xi} = 1$ may not be satisfied
\cite{Leinweber:1996ie,dblPiCloud,dblOctet,dblMagMomSR,dblDecuplet,dblDiquarks,Leinweber:2001ui,Leinweber:1999nf}.
This observation should be contrasted with the usual quark model
assumption that the quark magnetic moment is an intrinsic property,
independent of the quark's environment.

In three-flavor QCD, $O_N$ contains sea-quark-loop contributions from
$u$, $d$ and $s$ quarks.  In the SU(3)-flavor limit ($m_u = m_d =
m_s$) the charges add to zero and hence the sum vanishes.  However,
the heavier strange quark mass leads to a non-zero result.  By
definition
\begin{eqnarray}
O_N &=& \frac{2}{3} \,{}^{\ell}G_M^u - \frac{1}{3} \,{}^{\ell}G_M^d -
\frac{1}{3} \,{}^{\ell}G_M^s \, , \\
&=& \frac{{}^{\ell}G_M^s}{3} \left ( \frac{1 -
{}^{\ell}R_d^s}{{}^{\ell}R_d^s } \right ) \, , 
\label{OGMs}
\end{eqnarray}
where
\begin{equation}
{}^{\ell}R_d^s \equiv \frac{{}^{\ell}G_M^s}{{}^{\ell}G_M^d} \, ,
\label{lRsd}
\end{equation}
and the leading superscript, $\ell$, reminds one that the
contributions are loop contributions.  Note that, in deriving
Eq.(\ref{OGMs}), we have set ${}^{\ell}G_M^u = {}^{\ell}G_M^d$,
corresponding to $m_u = m_d$ \cite{Leinweber:1996ie}. Earlier
estimates of $^{\ell}R_d^s$ were based on the constituent quark model,
however a more reliable approach is to estimate the loops using a
finite range regulator
\cite{Young:2002ib,Leinweber:2003dg,Cloet:2003jm}.  Since the chiral
coefficients for the $d$ and $s$ loops in the RHS of Fig.~1 are
identical, the only difference comes from the mass of the $K$ compared
with that of the $\pi$.

With no more than a little accounting, the strange-quark loop
contributions to the nucleon magnetic moment, $G_M^s$ may be isolated
from (\ref{equalities}) and (\ref{OGMs}) in the following two
phenomenologically useful forms,
\begin{equation}
G_M^s = \left ( {\,{}^{\ell}R_d^s \over 1 - \,{}^{\ell}R_d^s }
\right ) \left [ 2 p + n - {u^p \over u^{\Sigma}} \left ( \Sigma^+ -
\Sigma^- \right ) \right ] ,
\label{GMsSigma}
\end{equation}
and
\begin{equation}
G_M^s = \left ( {\,{}^{\ell}R_d^s \over 1 - \,{}^{\ell}R_d^s } \right ) \left [
p + 2n - {u^n \over u^{\Xi}} \left ( \Xi^0 - \Xi^- \right ) 
 \right ] .
\label{GMsXi}
\end{equation}
As we have explained, under the assumption that quark magnetic
moments are not environment dependent, these ratios (i.e.
${u^p}/{u^{\Sigma}}$ and ${u^n}/{u^{\Xi}}$) are taken to be unity in
many quark models.  Incorporating the experimentally measured baryon
moments leads to:
\begin{equation}
G_M^s = \left ( {\,{}^{\ell}R_d^s \over 1 - \,{}^{\ell}R_d^s } \right ) \left [
3.673 - {u^p \over u^{\Sigma}} \left ( 3.618 \right ) \right ] , 
\label{ok}
\end{equation}
and
\begin{equation}
G_M^s = \left ( {\,{}^{\ell}R_d^s \over 1 - \,{}^{\ell}R_d^s } \right ) \left [
-1.033 - {u^n \over u^{\Xi}} \left ( -0.599 \right ) \right ] ,
\label{great}
\end{equation}
where all moments are expressed in nuclear magnetons $(\mu_N)$. (Note
that the measured magnetic moments are all known sufficiently
accurately \cite{PDG} that the experimental errors play no role in our
subsequent analysis.)  We stress that {\em these expressions for
$G_M^s$ are exact consequences of QCD, under the assumption of charge
symmetry}. Equation (\ref{great}) provides a particularly favorable
case for the determination of $G_M^s$ with minimal dependence on the
valence-quark ratio.

If one considers the quark model suggestions of $u^n/u^{\Xi}=1$ and
${}^{\ell}R_d^s=0.65$ in (\ref{great}), one finds $G_M^s = -0.81\
\mu_N$, a significant departure from the experimental preference of
positive values.

\begin{figure}[tbp]
\begin{center}
{\includegraphics[height=\hsize,angle=90]{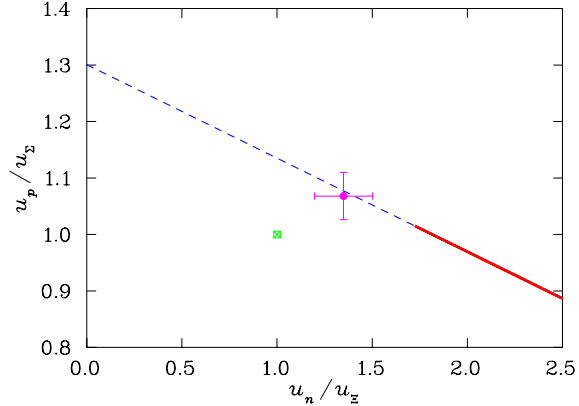}}
\end{center}
\vspace*{-1.0cm}
\caption{The consistency relation between $u^p/u^{\Sigma}$ and
$u^n/u^{\Xi}$ which must be satisfied within QCD.  The part of the
straight line which is dashed corresponds to $G_M^s(0) < 0$, while the
solid part of the line has $G_M^s(0) > 0$.  The standard quark model
assumption of intrinsic quark moments independent of their environment
is indicated by the filled square.  The lattice QCD prediction (after
an appropriate chiral extrapolation, discussed in following sections)
is illustrated by the filled circle.  
}
\label{SelfCons}
\end{figure}

Equating (\ref{ok}) and (\ref{great}) provides a linear relationship
between $u^p/u^{\Sigma}$ and $u^n/u^{\Xi}$ which must be satisfied
within QCD. Figure \ref{SelfCons} displays this relationship by the
dashed and solid line, the latter corresponding to values for which
$G_M^s(0) > 0$ when ${}^{\ell}R_d^s$ is in the anticipated range $0 <
{}^{\ell}R_d^s < 1$.  Since the line does not pass through the point
$(1.0, 1.0)$ corresponding to the simple quark model assumption of
universality, the experimentally measured baryon moments are signaling
that there must be an environment effect exceeding 12\% in both ratios
or approaching 20\% or more in at least one of the ratios.  Moreover,
a positive value for $G_M^s(0)$ requires an environment sensitivity
exceeding 70\% in the $u^n/u^{\Xi}$ ratio.  Hence the experimental
suggestion that $G_M^s(0) > 0$ challenges the intrinsic magnetic
moment concept which is fundamental to the constituent quark model.

\section{NUMERICAL SIMULATIONS WITH FLIC FERMIONS}

Access to leading-edge supercomputing resources coupled with advances
in the formulation of computationally-inexpensive chirally-improved
lattice fermion actions enable the numerical calculation of hadron
structure near the chiral regime.  The numerical simulations of the
electromagnetic form factors presented here are carried out using the
Fat Link Irrelevant Clover (FLIC) fermion action
\cite{Zanotti:2001yb,Leinweber:2002bw} in which the irrelevant
operators introduced to remove fermion doublers and lattice spacing
artifacts are constructed with smoothed links.  These links are
created via APE smearing \cite{ape}; a process that averages a link
with its nearest transverse neighbors in a gauge invariant manner.
Iteration of the averaging process generates a ``fat'' link.

The use of links in which short-distance fluctuations have been
removed simplifies the determination of the coefficients of the
improvement terms in both the action and its associated conserved
vector current.  Perturbative renormalizations are small for smeared
links and the mean-field improved coefficients used here are
sufficient to remove ${\mathcal O}(a)$ errors, in the lattice spacing
$a$, from the lattice fermion action.
The key is that both the energy dimension-five Wilson and Clover terms
\cite{Bilson-Thompson:2002jk} are constructed with smooth links, while
the relevant operators, surviving in the continuum limit, are
constructed with the original untouched links generated via standard
Monte Carlo techniques.

FLIC fermions provide a new form of nonperturbative ${\mathcal O}(a)$
improvement \cite{Leinweber:2002bw,inPrep} where near-continuum
results are obtained at finite lattice spacing.  Access to the light
quark mass regime is enabled by the improved chiral properties of the
lattice fermion action.  The magnitude of additive mass
renormalizations is suppressed \cite{inPrep} which otherwise can
lead to singular behavior in the propagators as the quarks become
light.

The ${\mathcal O}(a)$-improved conserved vector current
\cite{Martinelli:ny} is used.  Nonperturbative improvement is achieved
via the FLIC procedure where the terms of the Noether current having
their origin in the irrelevant operators of the fermion action are
constructed with mean-field improved APE smeared links.  The
preliminary results presented here are from a sample of 255 $20^3
\times 40$ mean-field improved Luscher-Weisz \cite{Luscher:1984xn}
gauge field configurations having a lattice spacing of 0.128 fm as
determined by the Sommer scale $r_0=0.50$ fm.

\section{CHIRAL EXTRAPOLATION}

One of the major challenges at present in connecting lattice
calculations of hadronic properties with the physical world is that
computational limitations restrict the accessible quark masses to
values much larger than the physical values.  At quark masses typical
of today's lattice QCD simulations, one is outside the region where
traditional dimensionally regulated (DR) chiral perturbation theory
($\chi$PT) is applicable.  Yet one knows that for current quark masses
near zero there is important non-analytic structure (as a function of
the quark mass) which must be treated correctly if we are to compare
with physical hadron properties.

Our present analysis of the strangeness magnetic form factor has been
made possible by a significant breakthrough in the regularization of
the chiral loop contributions to hadron observables
\cite{Young:2002ib,Leinweber:2003dg}.  Through the process of
regulating loop integrals via a finite-range regulator
\cite{Young:2002ib}, the chiral expansion is re-summed producing an
expansion with vastly improved convergence properties.

The key feature of finite-range regularization (FRR) is that FRR
schemes have an additional adjustable regulator parameter which
provides an opportunity to suppress short distance physics from the
loop integrals of effective field theory.  This short-distance
physics is otherwise treated incorrectly as the naive effective
fields do not share the properties of QCD at short distances.

An extensive study of the quark mass dependence of the nucleon mass in
finite-range regularized (FRR) \chiPT \cite{Young:2002ib} explored six
different regularization and associated renormalization schemes.  The
smooth regulators (dipole, monopole, or Gaussian regulators) provide
expansions that agree over the range $0 \le m_\pi^2 < \sim 0.8\ {\rm
  GeV}^2$ at a level sufficient to predict the nucleon mass within
1\%.  Our focus here is to extrapolate FLIC fermion calculations of
valence quark contributions to baryon moments $(u^p,\ u^n,\ 
u^{\Sigma},\ u^{\Xi})$ to the physical mass regime.  We select the
dipole-vertex FRR with $\Lambda = 0.8$ GeV as determined in
Ref.~\cite{Young:2002cj}. This scale was found to give the best
simultaneous description of both quenched and dynamical simulation
results.

Separation of the valence and sea-quark-loop contributions to the
meson cloud of full QCD hadrons is a non-trivial task.  We use the
diagrammatic method for evaluating the quenched chiral coefficients of
leading nonanalytic terms in heavy-baryon quenched \chiPT\
\cite{Leinweber:2001jc,Leinweber:2002qb}.  These results are
generalized to the FRR approach used here.  The diagrammatic technique
provides a transparent means to accomplish the separation of magnetic
moment contributions in full-QCD into ``direct sea-quark loop'' and
``valence'' contributions as in the discussion surrounding
Fig.~\ref{topology}.  The valence contributions (key to this analysis)
are obtained by removing the direct-current coupling to sea-quark
loops from the total contributions.  Upon further removing ``indirect
sea-quark loop'' contributions, where a valence quark forms a meson
composed with a sea-quark loop, one obtains the ``quenched valence''
contributions, the conventional view of the quenched approximation.
Isolating a particular quark flavour contribution only requires
setting the electric charge of all other quark flavours to zero.

\begin{figure}[tbp]
\begin{center}
  {\includegraphics[height=\hsize,angle=90]{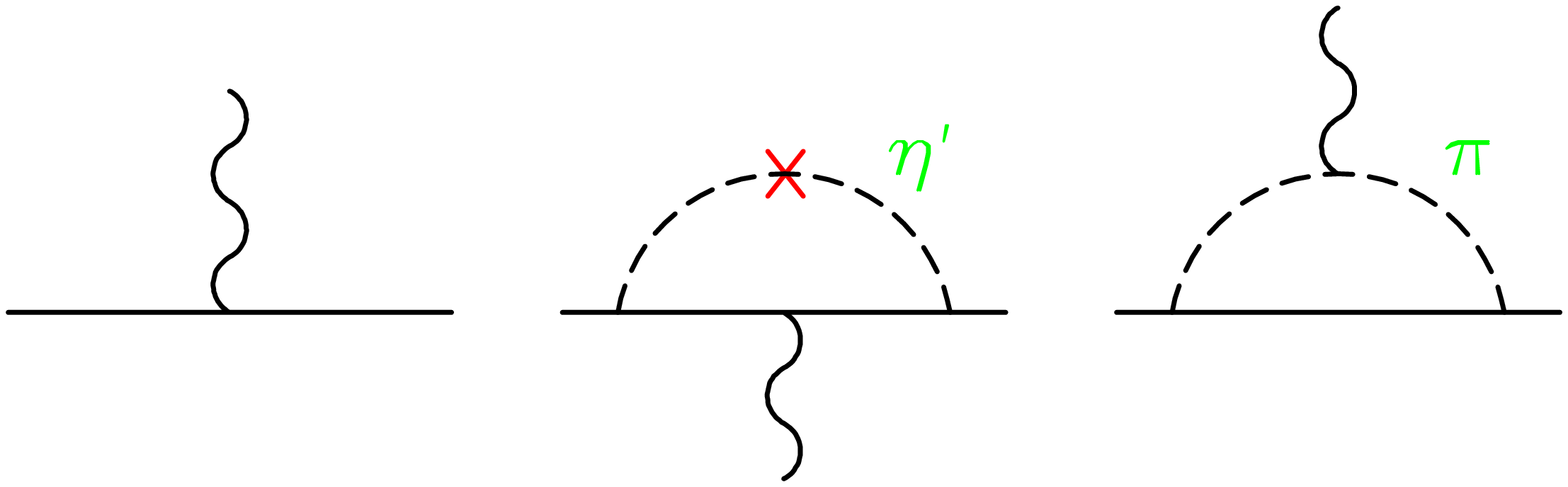}}

  {\includegraphics[height=\hsize,angle=90]{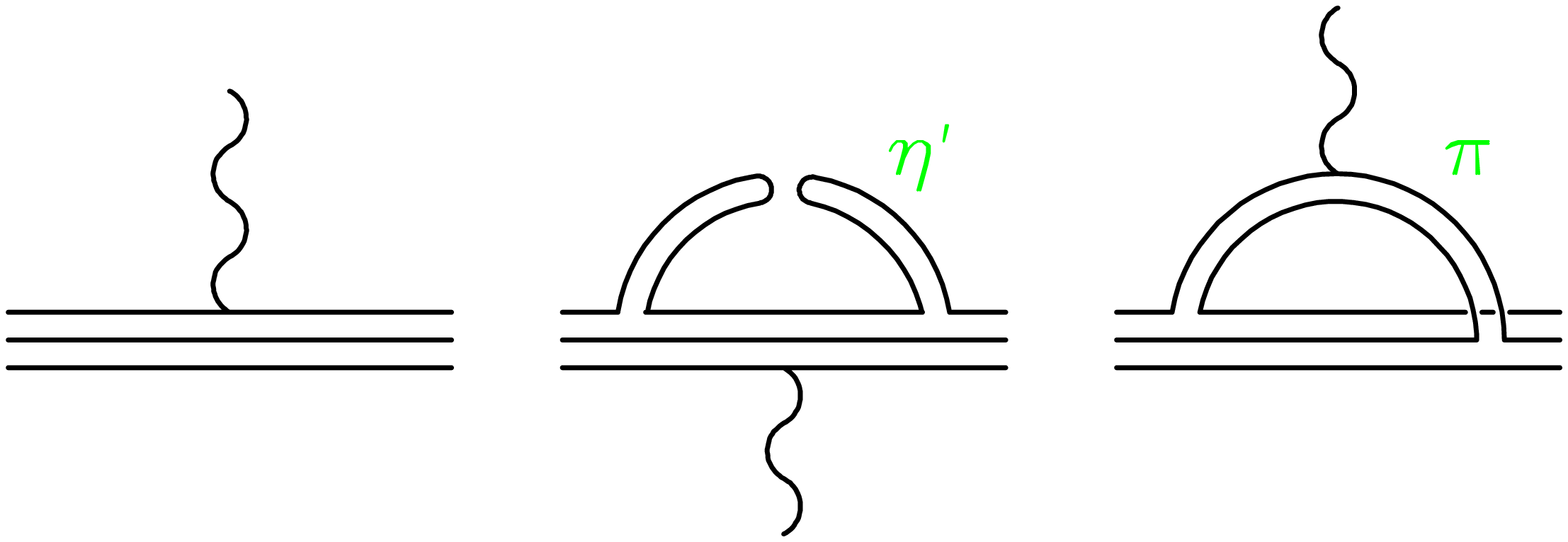}}
\end{center}
\vspace*{-1.0cm}
\caption{
Diagrams providing the leading contributions to the chiral expansion
of baryon magnetic moments (upper diagrams) and their associated quark
flows (lower diagrams) in QQCD.  In the lower diagrams, the
quark-photon coupling is summed over adjacent quark lines.
}
\label{QquarkFlow}
\end{figure}

Figure~\ref{QquarkFlow} displays the diagrams providing the leading
contributions to the chiral expansion of baryon magnetic moments
(upper diagrams) and their associated quark flows in quenched QCD
(QQCD).  The associated chiral expansion for the proton magnetic
moment, $\mu_p$, has the form
\begin{eqnarray}
\mu_p &=& a_0^\Lambda 
            + \mu_p \, \chi_{\eta'} \, I_{\eta'}(m_\pi, \Lambda)
            + \chi_\pi \, I_\pi(m_\pi, \Lambda) \nonumber \\
       &&   + \chi_K \, I_K(m_K, \Lambda)
            + a_2^\Lambda \, m_\pi^2 
            + a_4^\Lambda \, m_\pi^4 \, .
\label{chiExp}
\end{eqnarray}
where $I$ denotes a loop integral defined by
{\small
\begin{eqnarray}
&&\hspace*{-5mm}I_\pi(m_\pi,\Lambda) = -\frac{4}{3\pi}\int_0^\infty dk\,
\frac{k^4}{(k^2+m_\pi^2)^2}u^2(k,\Lambda) \\
&&\hspace*{-5mm}I_K(m_K, \Lambda)    = \nonumber\\&&\hspace*{-2mm}-\frac{4}{3\,\pi}\int dk \frac{k^4
  u^2(k)}{(k^2+m_K^2)(\sqrt{k^2+m_K^2} + \Delta_{BN})^2} \\
&&\hspace*{-5mm}I_{\eta'}(m_\pi, \Lambda) = - \int_0^\infty dk\,
\frac{k^4}{(k^2+m_\pi^2)^\frac{5}{2}}u^2(k,\Lambda)
\end{eqnarray}
}
where $\Delta_{BN}$ is the physical mass-splitting between the $N$ and
$\Sigma$ or $\Lambda$.  The coefficients, $\chi$, denote the
model-independent coefficients of the LNA term for $\pi$ and $K$
mesons \cite{Leinweber:2002qb}.  We take
\begin{equation}
m_K^2 = m_K^{(0)\, 2} + \frac{1}{2} \, m_\pi^2
\end{equation}
where the physical values may be used to define $ m_K^{(0)}$.  The
$m_\pi^4$ term of Eq.~(\ref{chiExp}) allows for some curvature
associated with the Dirac moment of the baryon $\propto 1/m_\pi^2$ for
moderately large quark masses.

The un-renormalized coefficients of the analytic terms of the FRR
expansion are regulator-parameter dependent.  This is emphasized by
the superscript $\Lambda$ on the coefficients $a_0,\ a_2$ and $a_4$.
The large $m_\pi$ behavior of the loop integrals of Eq.~(\ref{chiExp})
and the residual expansion are remarkably different.  Whereas the
residual expansion will encounter a power divergence, the FRR loop
integrals will tend to zero as some power of $\Lambda/m_\pi$, as
$m_\pi$ becomes large.  Thus, the $\Lambda$ dependence of $a_0,\ a_2$
and $a_4$ provides an opportunity to govern the convergence properties
of the residual expansion and thus the FRR chiral expansion.

Lattice QCD has now provided extensive model-independent information
on the moderate to large $m_\pi$ dependence of hadron observables.  In
particular, hadron masses are observed to be smooth almost linear
functions of $m_\pi^2$ for quark masses similar to the strange quark
mass.  The magnetic moments presented here are also smooth, taking on
a Dirac moment dependence as the quark mass becomes large.  This
indicates that it should be possible to tune the regulator-range
parameter, $\Lambda$, such that the coefficients $a_4$, and higher are
truly small.  In this case the convergence properties of the residual
expansion, and the loop expansion are excellent and their truncation
benign.  Indeed this hypothesis was confirmed for the nucleon mass in
Ref.~\cite{Young:2002ib}.

It is important to note that the optimal $\Lambda$ is one that
optimizes the convergence properties of the residual expansion.  A
poor choice for $\Lambda$ will move strength in terms proportional to
1, $m_\pi^2$, $m_\pi^4,\ldots$ from the loop integrals into the
residual expansion, changing the convergence properties of the FRR
expansion.  We emphasize that the optimal $\Lambda$ is {\it not}
selected to approximate the higher-order terms of the chiral
expansion.  If an infinite number of parameters were being approximated
by one, one would have an uncontrolled error.  Fortunately such ambitions are
not necessary.  The lattice QCD results indicate that these higher
order terms of the chiral expansion largely sum to zero.  Moreover,
the findings of Ref.~\cite{Young:2002ib} indicate that the details of
exactly how each of the terms enters the vanishing sum are largely
irrelevant.
 
Perhaps it is also worth emphasizing that the choice of $\Lambda$ can
have no effect on the convergence properties of the renormalized
chiral expansion.  There, contributions from the residual expansion
and the FRR loop integrals are combined, removing any $\Lambda$
dependence from the chiral expansion.  This is the process of
renormalization.

Donoghue {\it et al.} \cite{Donoghue:1998bs} describe the convergence
problems of DR as due to incorrect short-distance physics in loop
integrals which must be removed by large and opposing analytic-term
contributions.  The introduction of an ultraviolet cut off in FRR
schemes prevents the error from being made in the first place and
provides the opportunity to have excellent convergence properties in
the residual expansion.  In other words, the higher-order analytic
terms of the FRR residual expansion do not need to be large, as
incorrect short distance physics has been suppressed from the loop
integrals.  The beauty of the FRR procedure is that upon
renormalization, very large coefficients can be recovered
\cite{Leinweber:2003dg}.

\begin{figure}[tbp]
\begin{center}
{\includegraphics[height=\hsize,angle=90]{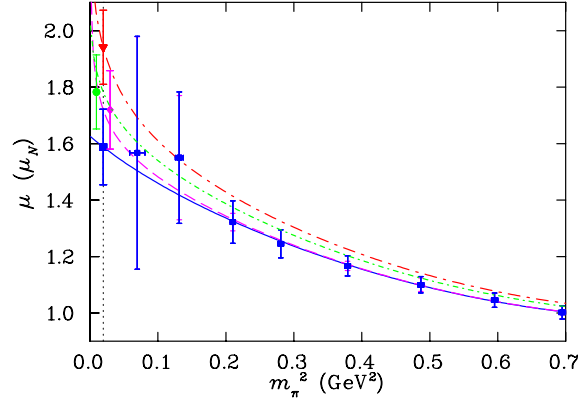}}
\end{center}
\vspace*{-1.0cm}
\caption{The contribution of a single $u$ quark normalized to unit
  charge to the magnetic moment of the proton.  Lattice simulation
  results (square symbols for $m_\pi^2 > 0.05$ GeV) are extrapolated
  to the physical point (vertical dashed line) in finite-volume QQCD
  with a discrete momentum sum (solid curve, square symbol), and in
  infinite-volume QQCD (long-dashed curve, diamond symbol).  Estimates
  of the valence $u$ quark contribution in full QCD (dot-dash curve,
  triangular symbol) and the full $u$ quark sector contribution in
  full QCD (fine-dash curve, circular symbol) normalized to a single
  quark unit charge are also illustrated.  Extrapolated values
  indicated by symbols at the physical pion mass (vertical dashed
  line) are offset for clarity. }
\label{uProton}
\end{figure}

Figure \ref{uProton} illustrates a fit of FRR Q\chiPT to the FLIC
fermion lattice results (solid curve), in a calculation where only the
discrete momenta allowed in the finite volume of the lattice are
summed in performing the loop integral.  The long-dashed curve that
also runs through the lattice results illustrates the case when the
discrete momentum sum is replaced by the infinite-volume, continuous
momentum integral.  For all but the lightest quark mass, finite volume
effects are negligible.

The coefficients of the residual expansion, $a_0^\Lambda,\ a_2^\Lambda,\
a_4^\Lambda$, show excellent signs of convergence with values 1.52(14),
$-$1.23(48), and 0.69(42) in appropriate powers of GeV, respectively.
 
Incorporating baryon mass splittings into the kaon loop contributions
is essential.  For example the contribution of $\Sigma \to N K$ is
nearly doubled when the $\Sigma - N$ mass splitting is taken into
account. 

\begin{figure}[tbp]
\begin{center}
{\includegraphics[height=\hsize,angle=90]{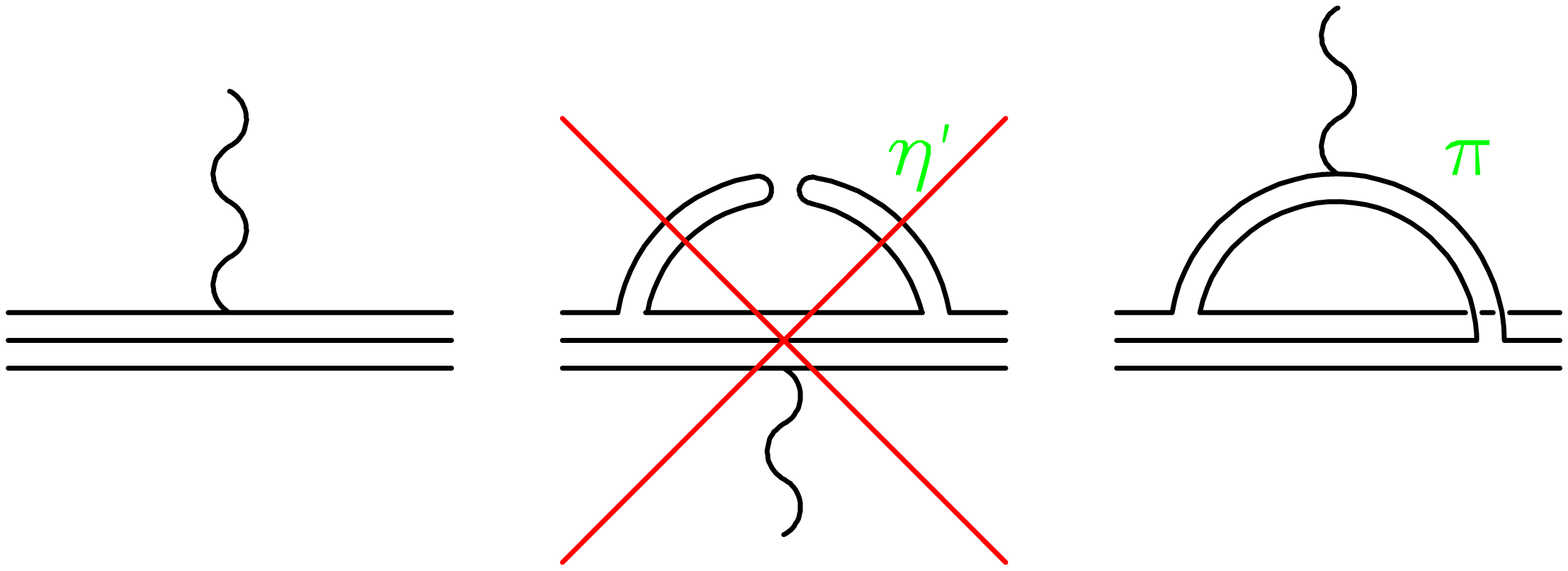}}

{\includegraphics[height=\hsize,angle=90]{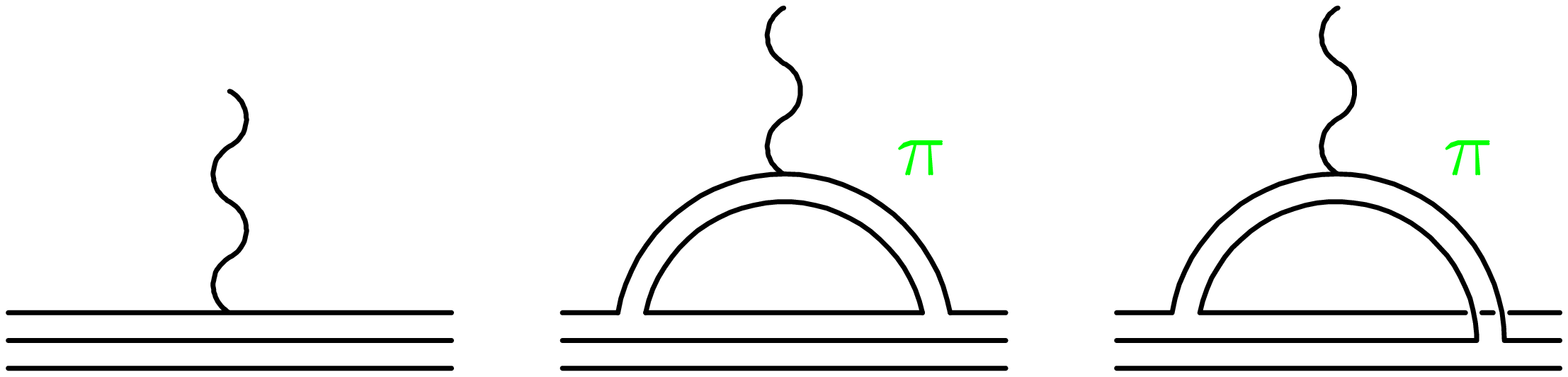}}
\end{center}
\vspace*{-1.0cm}
\caption{
Correcting Q\chiPT (upper diagrams) to full QCD (lower diagrams).
Quenched negative-metric $\eta'$ contributions must be removed, while
the chiral coefficients of $\pi$ and $K$ loops must be adjusted to
account for the new diagram where a valence quark couples to the
photon in a meson constructed from a sea-quark loop.  Photon couplings
to the anti-quark in the bottom-right diagram are also included in the
valence quark contributions to the nucleon moment of full QCD.
}
\label{unquenching}
\end{figure}

Figure \ref{unquenching} diagrammatically illustrates the
considerations in correcting the quenched $u$-quark contribution to
the valence $u$-quark contribution in full QCD.  The removal of
quenched $\eta'$ contributions and the appropriate adjustment of $\pi$
and $K$ loop coefficients as detailed in
Ref.~\cite{Leinweber:2001jc,Leinweber:2002qb} provides the dot-dash
curve of Fig.~\ref{uProton}.  This is our best estimate of the valence
$u$-quark contribution (the connected insertion of the current) to the
proton magnetic moment of full QCD.  The graph corresponds to a single
$u$-quark contribution normalized to unit charge.  It will be interesting to
confront this curve with full QCD simulation results in the future.

\begin{figure}[tbp]
\begin{center}
{\includegraphics[height=\hsize,angle=90]{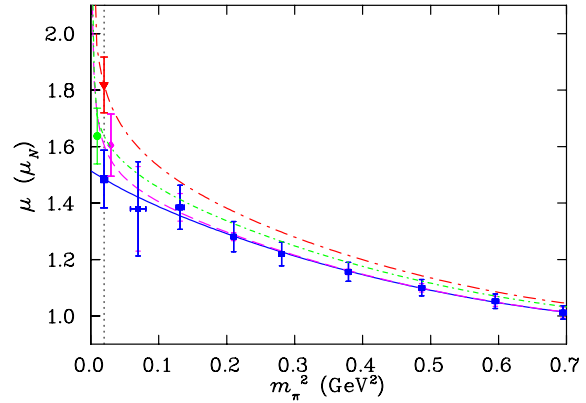}}
\end{center}
\vspace*{-1.0cm}
\caption{
The contribution of a single $u$ quark normalized to unit
charge to the magnetic moment of $\Sigma^+$.  Curves and symbols are
as described in Fig.~\protect\ref{uProton}.
}
\label{uSigma}
\end{figure}

\begin{figure}[tbp]
\begin{center}
{\includegraphics[height=\hsize,angle=90]{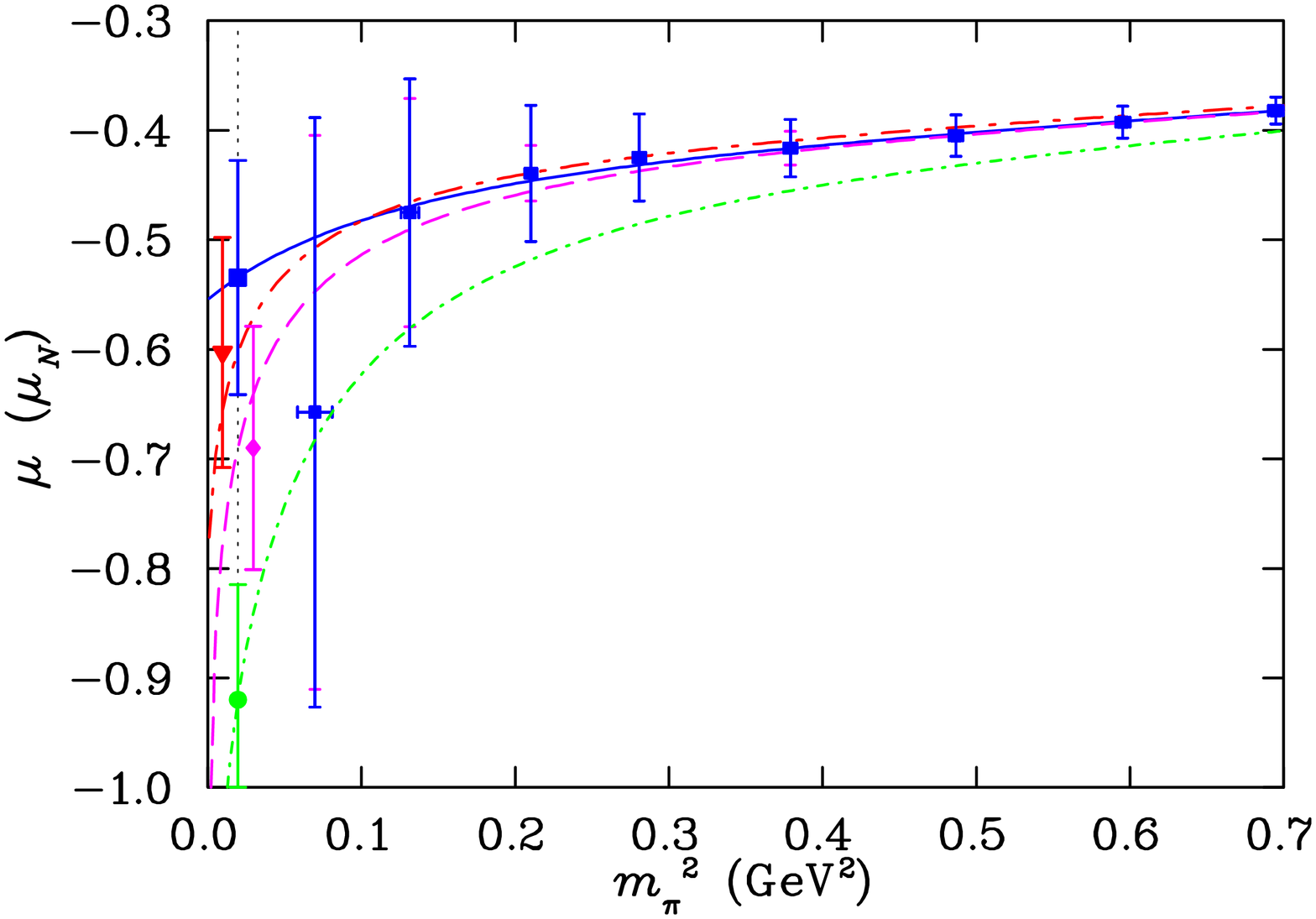}}
\end{center}
\vspace*{-1.0cm}
\caption{ 
The contribution of the $u$ quark normalized to unit charge
to the magnetic moment of the neutron.  Curves and symbols are as
described in Fig.~\protect\ref{uProton}.
}
\label{uNeutron}
\end{figure}

\begin{figure}[tbp]
\begin{center}
{\includegraphics[height=\hsize,angle=90]{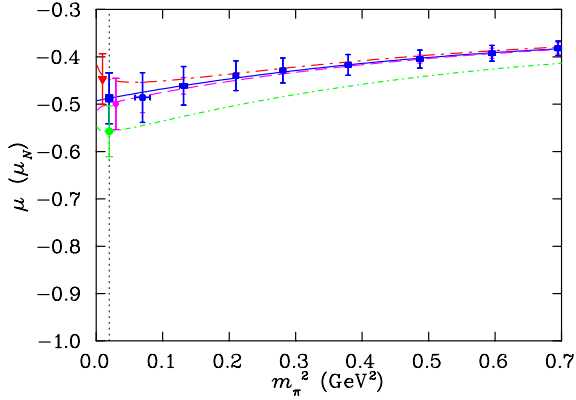}}
\end{center}
\vspace*{-1.0cm}
\caption{
The contribution of the $u$ quark normalized to unit charge
to the magnetic moment of $\Xi^0$.  Curves and symbols are as
described in Fig.~\protect\ref{uProton}.
}
\label{uXi}
\end{figure}

Finally we also include the disconnected insertion of the current in
order to estimate the total contribution of the $u$-quark sector to
the proton magnetic moment.  This is represented by the fine dashed
curve in Fig.~\ref{uProton} where the sector contribution is
normalized to a single quark of unit charge. Figures \ref{uSigma},
\ref{uNeutron} and \ref{uXi} display similar results for the
$\Sigma^+$, $n$ and $\Xi^0$.

From these chiral extrapolations, we estimate the ratio of the valence
(connected) $u$-quark contribution to the proton to that in
$\Sigma^+$.  Similarly, the ratio of the $u$-quark contribution to the
neutron relative to that in $\Xi^0$ is calculated.  Error bars are
estimated from a third-order single-elimination jackknife analysis of
the correlated ratios.  Our final result of 
\begin{equation}
\frac{u^p}{u^\Sigma} = 1.07\pm 0.04
\quad
\mbox{and}
\quad
\frac{u^n}{u^\Xi} = 1.35\pm 0.15
\label{uRatios}
\end{equation}
is plotted in Fig.~\ref{SelfCons}.  This result leaves little doubt
that $G_M^s$ is negative.  The fact that this point lies exactly on the
constraint curve is highly nontrivial, and provides a robust check on
the validity of the analysis techniques presented here.

\begin{figure}[tbp]
\begin{center}
{\includegraphics[height=\hsize,angle=90]{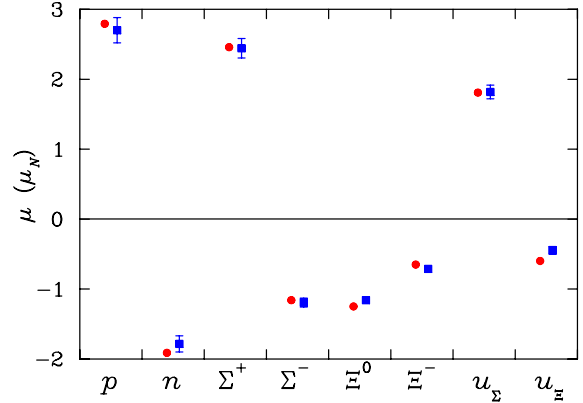}}
\end{center}
\vspace*{-1.0cm}
\caption{
Comparison of the \FRRchiPT extrapolated lattice QCD simulation
results (square symbols) with experimentally measured baryon magnetic
moments (circular symbols).
}
\label{compExpt}
\end{figure}

As a further check we compare the lattice QCD predictions of the
baryon magnetic moments constructed from extrapolations of the
individual quark sectors in Fig.~\ref{compExpt}.  The results display
an unprecedented agreement with experiment.  There is a hint of a
small systematic error that suppresses the singly represented quark
contribution to both $\Xi$ baryons and the neutron.  We have
identified contributions from decuplet baryons as key to resolving
this minor discrepancy.  However, the ratio of $u^n/u^\Xi$ does not
appear to be affected by this contribution, as evidenced by the
excellent agreement between lattice and experiment for the $u$ quark
in $\Sigma$ and the degree of consistency between the lattice
extrapolations and the constraint curve of Fig.~\ref{SelfCons}.
The fact that systematic errors will cancel in taking the ratios of
specific quark contributions was the main motivation for introducing the
ratios in the first place.

While $G_M^s$ is most certainly negative, it remains to set the
magnitude.  This requires an estimate of the strange to light
sea-quark loop contributions, ${}^{\ell}R_d^s$.  Earlier estimates of
$^{\ell}R_d^s$ were based on the constituent quark model, however a
more reliable approach is to estimate the loops using a finite range
regulator \cite{Young:2002ib,Leinweber:2003dg}.  Using a  
dipole-vertex regulator, a
direct calculation provides 
\begin{equation} 
{}^{\ell}R_d^s = \frac{G_M^s}{G_M^d} = 
  0.160 \pm 0.060 
\, , 
\label{Rsd}
\end{equation} 
where the range includes the variation of the dipole mass parameter
between 0.6 and 1.0 GeV as well as the possible variation of the mass of
the strange baryon accompanying the $K^+$ loop.
Using this value in Eq.~(\ref{GMsXi}) with the results of
Eq.~(\ref{uRatios}) provides the final precise determination of
\begin{equation} 
{G_M^s} = -0.043 \pm 0.026\ \mu_N \, ,
\label{GMs}
\end{equation} 
for the strange quark contribution to the magnetic moment of the
nucleon.  This extremely precise value sets a tremendous challenge for
the next generation of parity violation experiments.

\section*{Acknowledgments}
We thank the Australian Partnership for Advanced Computing (APAC) for
generous grants of supercomputer time which have enabled this project.
This work is supported by the Australian Research Council.


\begin{thebibliography}{99}

\bibitem{Aniol:2000at}
K.~A.~Aniol {\it et al.}  [HAPPEX Collaboration],
Phys.\ Lett.\ B {\bf 509}, 211 (2001)
[arXiv:nucl-ex/0006002].

\bibitem{Hasty:2001ep}
R.~Hasty {\it et al.}  [SAMPLE Collaboration],
Science {\bf 290}, 2117 (2000)
[arXiv:nucl-ex/0102001].

\bibitem{Mathur:2000cf}
N.~Mathur and S.~J.~Dong  [Kentucky Field Theory Collaboration],
Nucl.\ Phys.\ Proc.\ Suppl.\  {\bf 94} (2001) 311
[arXiv:hep-lat/0011015].

\bibitem{Lewis:2002ix}
R.~Lewis, W.~Wilcox and R.~M.~Woloshyn,
Phys.\ Rev.\ D {\bf 67} (2003) 013003
[arXiv:hep-ph/0210064].

\bibitem{ChargeSymm}
G.A.~Miller, B.M.~Nefkens and I.~Slaus,
Phys.\ Rept.\ {\bf 194} (1990) 1.

\bibitem{Leinweber:1996ie}
D.B.~Leinweber,
Phys.\ Rev.\ {\bf D53}, 5115 (1996)
hep-ph/9512319.

\bibitem{dblPiCloud}
T.~D. Cohen and D.~B. Leinweber.
Comments Nucl.\ Part.\ Phys.\ {\bf 21}, 137 (1993) hep-ph/9212225;
A.W.~Thomas,
Austral.\ J.\ Phys.\ {\bf 44}, 173 (1991).
 
\bibitem{dblOctet}
D.~B. Leinweber, R.~M. Woloshyn, and T.~Draper,
Phys.\ Rev.\ {\bf D43}, 1659 (1991).

\bibitem{dblMagMomSR}
D.~B. Leinweber,
Phys.\ Rev.\ {\bf D45}, 252 (1992).

\bibitem{dblDecuplet}
D.~B. Leinweber, T.~Draper, and R.~M. Woloshyn.
Phys.\ Rev.\ {\bf D46}, 3067 (1992) hep-lat/9208025.

\bibitem{dblDiquarks}
D.~B. Leinweber,
Phys.\ Rev.\ {\bf D47}, 5096 (1993) hep-ph/9302266.

\bibitem{Leinweber:2001ui}
D.~B.~Leinweber, A.~W.~Thomas and R.~D.~Young,
Phys.\ Rev.\ Lett.\  {\bf 86}, 5011 (2001)
[arXiv:hep-ph/0101211].
%
\bibitem{Leinweber:1999nf}
D.~B.~Leinweber and A.~W.~Thomas,
Phys.\ Rev.\ D {\bf 62} (2000) 074505
[arXiv:hep-lat/9912052].
%
\bibitem{Young:2002ib}
R.~D.~Young {\it et al.},
Prog.\ Part.\ Nucl.\ Phys.\  {\bf 50} (2003) 399
[arXiv:hep-lat/0212031].

\bibitem{Leinweber:2003dg}
D.~B.~Leinweber, A.~W.~Thomas and R.~D.~Young,
arXiv:hep-lat/0302020.

\bibitem{Cloet:2003jm}
I.~C.~Cloet, D.~B.~Leinweber and A.~W.~Thomas,
Phys.\ Lett.\ B {\bf 563} (2003) 157
[arXiv:hep-lat/0302008].

\bibitem{PDG}
Particle Data Group, Phys.\ Rev.\ {\bf D66} (2002) 010001.

\bibitem{Zanotti:2001yb}
J.~M.~Zanotti {\it et al.}  [CSSM Lattice Collaboration],
Phys.\ Rev.\ D {\bf 65} (2002) 074507
[arXiv:hep-lat/0110216].

\bibitem{Leinweber:2002bw}
D.~B.~Leinweber {\it et al.},
arXiv:nucl-th/0211014.

\bibitem{ape}
M.~Falcioni {\it et al.}, 
Nucl.\ Phys.\ {\bf B251}, 624 (1985).

\bibitem{Bilson-Thompson:2002jk}
S.~O.~Bilson-Thompson {\it et al.},
Annals Phys.\  {\bf 304} (2003) 1
[arXiv:hep-lat/0203008].

\bibitem{inPrep}
J.~Zanotti, {\it et al.}, in preparation.

\bibitem{Martinelli:ny}
G.~Martinelli, C.~T.~Sachrajda and A.~Vladikas,
Nucl.\ Phys.\ B {\bf 358} (1991) 212.

\bibitem{Luscher:1984xn}
M.~Luscher and P.~Weisz,
Commun.\ Math.\ Phys.\  {\bf 97}, 59 (1985)
[ibid.\  {\bf 98}, 433 (1985)].

\bibitem{Young:2002cj}
R.~D.~Young, D.~B.~Leinweber, A.~W.~Thomas and S.~V.~Wright,
Phys.\ Rev.\ D {\bf 66}, 094507 (2002)
[arXiv:hep-lat/0205017].

\bibitem{Leinweber:2001jc}
D.~B.~Leinweber,
Nucl.\ Phys.\ Proc.\ Suppl.\  {\bf 109A} (2002) 45
[arXiv:hep-lat/0112021].

\bibitem{Leinweber:2002qb}
D.~B.~Leinweber,
arXiv:hep-lat/0211017.

\bibitem{Donoghue:1998bs}
J.~F. Donoghue, B.~R. Holstein and B.~Borasoy,
\newblock Phys. Rev. {\bf D59}, 036002 (1999).


\end{thebibliography}
\end{document}